\documentclass[amsmath, preprint, floatfix]{revtex4-2}
\usepackage{graphicx}
\usepackage{microtype}
\usepackage{bm}
\usepackage{physics}
\usepackage[detect-weight=true, detect-family=true]{siunitx}

\begin{document}

\title{Soliton self-compression and resonant dispersive wave emission in higher-order modes of a hollow capillary fibre}

\author{Christian Brahms}
\email[Corresponding author: ]{c.brahms@hw.ac.uk}
\author{John C. Travers}
\affiliation{School of Engineering and Physical Sciences, Heriot-Watt University, Edinburgh, EH14 4AS, UK}

\begin{abstract}
We investigate soliton self-compression and ultraviolet resonant dispersive wave emission in the higher-order modes of a gas-filled hollow capillary fibre. Our simple analytical scaling rules predict shorter required waveguides and different energy scales when moving from the fundamental to higher-order modes. Experimentally, we demonstrate soliton self-compression and ultraviolet dispersive wave emission in the double-lobe LP$_{11}$ mode of an argon-filled hollow capillary fibre, which we excite by coupling into the fibre at oblique incidence. We observe the generation of ultraviolet dispersive waves which are frequency-shifted and more narrowband as compared to fundamental-mode generation due to the stronger modal dispersion, and a suppression of the supercontinuum between the dispersive wave and the pump pulse. With numerical simulations, we confirm the predictions of our scaling rules and find that the use of higher-order modes can suppress photoionisation and plasma effects even while allowing for much higher pulse energy to be used in the self-compression process. Our results add another degree of freedom for the design of hollow-waveguide systems to generate sub-cycle field transients and tuneable ultrashort laser pulses.
\end{abstract}

\maketitle
\section{Introduction}
Gas-filled hollow-core waveguides are a very flexible platform for nonlinear frequency conversion as well as temporal and spectral reshaping of ultrafast laser pulses~\cite{travers_ultrafast_2011,russell_hollow-core_2014,nagy_high-energy_2021}. Soliton self-compression and other associated dynamics in these systems offer some unique capabilities, such as pulse compression to sub-cycle field transients and frequency conversion to few-femtosecond pulses from the vacuum ultraviolet to the near infrared~\cite{joly_bright_2011,mak_tunable_2013,ermolov_supercontinuum_2015,travers_high-energy_2019,brahms_infrared_2020}. These effects were first demonstrated in hollow-core photonic crystal fibres, which offer low-loss guidance over very large bandwidths and strong anomalous dispersion~\cite{joly_bright_2011}. However, their small core size (few tens of \si{\micro\meter}) in combination with intensity limits due to nonlinear effects in the filling gas sets a maximum peak power which can be used to drive soliton dynamics, with typical pulse energies below \SI{20}{\micro\joule}~\cite{joly_bright_2011,mak_tunable_2013,ermolov_supercontinuum_2015,kottig_generation_2017}. Subsequent work proposed that soliton dynamics at higher energy could be achieved by propagating pulses in simple hollow capillary fibres (HCFs) with much larger cores and compensating the resulting reduction in anomalous waveguide dispersion by exploiting higher-order modes~\cite{lopez-zubieta_theoretical_2018,lopez-zubieta_spatiotemporal-dressed_2018}. It was later demonstrated that the full range of soliton dynamics can in fact be obtained in the fundamental mode of an HCF for correctly chosen parameters~\cite{travers_high-energy_2019}. However, the properties of higher-order modes offer an additional degree of freedom which has so far not been explored experimentally.

In this work, we investigate soliton dynamics in the higher-order modes of gas-filled hollow-core waveguides. We derive simple scaling rules for the changes in the required pulse energy and length scale which result from a change in mode order and we experimentally demonstrate soliton self-compression and deep ultraviolet resonant dispersive wave (RDW) emission in a higher-order mode of an argon-filled HCF. As a simple alternative to coupling schemes involving active beam shaping, we excite the LP$_{11}$-like double-lobe mode by focusing a Gaussian-like beam onto the HCF entrance face at an angle. Despite this relatively crude method and the resulting mode mixture at the beginning of the nonlinear propagation, we find that the mode-dependent phase-matching to dispersive waves leads to clean higher-order mode profiles for the generated DUV pulses. Comparing RDW emission in LP$_{11}$ and the fundamental mode, we find significant differences in the bandwidth of the RDW as well as in the supercontinuum between the RDW and the pump. Numerical simulations confirm that self-compression and RDW emission occur over a shorter length scale in LP$_{11}$ than in the fundamental mode and reveal that photoionisation can be strongly suppressed by propagating in other higher-order modes despite an increase in total pulse energy.

\section{Properties of solitons in higher-order modes}
\label{sec:properties}
The central features of soliton propagation are created by the interplay of the Kerr nonlinearity and anomalous group-velocity dispersion. To observe full soliton self-compression and resonant dispersive emission in a hollow capillary fibre, the parameters must additionally be chosen so as to avoid excessive propagation losses~\cite{travers_high-energy_2019}. The mode in which the pulse propagates affects all three of these parameters.

The frequency-dependent propagation constant $\beta_{nm}(\omega)$ and attenuation coefficient $\alpha_{nm}(\omega)$ in a gas-filled capillary are given by
\begin{equation}
    \beta_{nm}(\omega) = \frac{\omega}{c}\sqrt{n_\mathrm{gas}^2(\omega) - \frac{c^2u_{nm}^2}{a^2\omega^2}}\,,
    \quad
    \alpha_{nm}(\omega, z) = \frac{c^2 u_{nm}^2}{a^3\omega^2} \nu_n(\omega)\,,
\end{equation}
where $\omega$ is angular frequency, $c$ is the speed of light, $n_\mathrm{gas}$ is the refractive index of the filling gas, and $a$ is the core radius. $\nu_n$ depends on the refractive index ratio between cladding and core as well as the character of the mode---transverse electric (TE$_{0m}$), transverse magnetic (TM$_{0m}$) or hybrid (HE$_{nm}$)---and $u_{nm}$ is the $m^\mathrm{th}$ zero of the Bessel function of the first kind $J_1$ (for TE/TM) or $J_{n-1}$ (for HE)~\cite{marcatili_hollow_1964}. $u_{nm}$ is smallest for the fundamental mode (HE$_{11}$), $u_{11} \approx 2.405$. The mode character in combination with the azimuthal and radial mode orders $n$ and $m$ designate a specific mode of the HCF. In addition, the hybrid modes are degenerate with respect to their polarisation direction, with the two degenerate modes rotated by $\pi/2n$ relative to each other.

The group-velocity dispersion is given by
\begin{equation}
    \beta^{(2)}_{nm}(\omega) = \partial^2_\omega \beta_{nm}(\omega) \approx \frac{\rho}{c}\left[\partial_\omega\gamma_\mathrm{gas}(\omega) + \frac{\omega}{2}\partial_\omega^2\gamma_\mathrm{gas}(\omega)\right] - \frac{c u_{nm}^2}{a^2 \omega^3} = \rho\delta_\mathrm{gas}(\omega) - \frac{c u_{nm}^2}{a^2 \omega^3}\,,
    \label{eq:beta2}
\end{equation}
where $\rho$ is the number density of the gas, $\gamma_\mathrm{gas}(\omega)$ is the linear polarisability of a single gas particle such that $n_\mathrm{gas}^2(\omega) = 1+\rho\gamma_\mathrm{gas}(\omega)$, and hence $\rho\delta_\mathrm{gas}(\omega)$ is the dispersion of the gas. The final term in eq.~\ref{eq:beta2} contains the waveguide contribution to the dispersion, which is always anomalous. Moving from the fundamental mode to higher-order modes means $u_{nm}$ becomes larger, which increases this anomalous dispersion contribution in a similar fashion to decreasing the core radius $a$. For a fixed pump wavelength, the dispersion landscape can be parameterised by the zero-dispersion wavelength $\lambda_\mathrm{zd} = 2\pi c/\omega_\mathrm{zd}$, with $\omega_\mathrm{zd}$ determined by
\begin{equation}
    \beta^{(2)}_{nm}(\omega_\mathrm{zd}) = 0 \quad \iff \quad \rho_\mathrm{zd} = \frac{c u_{nm}^2}{a^2 \omega_\mathrm{zd}^3 \delta_\mathrm{gas}(\omega_\mathrm{zd})}\,.
\end{equation}
The gas density required to obtain a certain dispersion landscape and phase-match RDW emission at a certain wavelength is thus proportional to $u_{nm}^2$, conversely to the scaling with core size \cite{travers_high-energy_2019,heyl_scale-invariant_2016}. Importantly, at all frequencies other than $\omega_\mathrm{zd}$, the dispersion increases in magnitude when $u_{nm}$ is increased and the density is adjusted to keep $\omega_\mathrm{zd}$ fixed. Hence, the dispersion at the pump wavelength becomes more strongly anomalous for higher mode orders.

\begin{figure}
    \centering
    \includegraphics[width=6in]{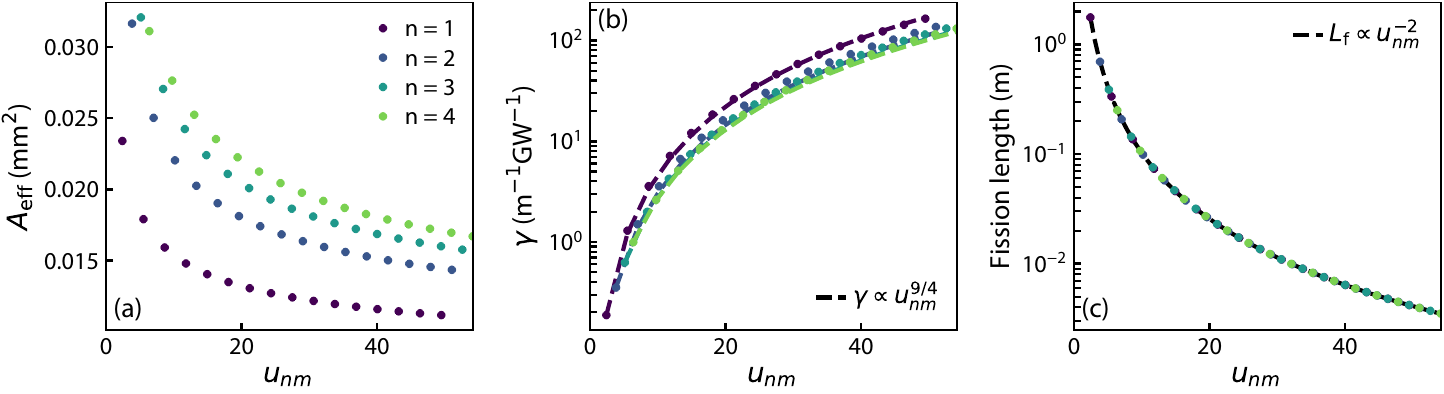}
    \caption{Change in the effective area (a), nonlinear coefficient (b) and fission length (c) with mode order for soliton self-compression in an argon-filled hollow capillary fibre with \SI{125}{\micro\meter} core radius. In (b) and (c), the zero-dispersion wavelength is held constant at $\lambda_\mathrm{zd} = \SI{500}{\nm}$. In (c), the soliton order is additionally held constant at $N=2.5$.}
    \label{fig:mode_properties}
\end{figure}
The nonlinearity of the gas-filled waveguide is captured by the nonlinear coefficient $\gamma = k_0 n_2 A_\mathrm{eff}^{-1}$, where $k_0=\omega/c$, $n_2$ is the nonlinear refractive index of the filling gas and $A_\mathrm{eff}$ is the effective area of the mode profile:
\begin{equation}
    A_\mathrm{eff} = \frac{
        \qty[
            \int_0^a \int_0^{2\pi} \abs{\bm{e}(r, \theta)}^2 \,r\dd r \dd \theta
        ]^2
    }
    {
        \int_0^a \int_0^{2\pi} \abs{\bm{e}(r, \theta)}^4 \,r\dd r \dd \theta
    }\,,
\end{equation}
where $\bm{e}(r, \theta)$ is the field distribution of the HCF mode in polar coordinates; bold symbols indicate vector quantities. The magnitude of $\bm{e}(r, \theta)$ for the modes of a hollow capillary fibre is given by $\abs{\bm{e}(r, \theta)} = J_i\qty(u_{nm}r/a)$, where $i = 1$ for TE/TM modes and $i=n-1$ for HE modes \cite{marcatili_hollow_1964}. The effective area increases for larger azimuthal indices $n$ and decreases for larger radial indices $m$ as shown in Fig.~\ref{fig:mode_properties}(a). When matching the dispersion, however, the increase in gas density dominates over the relatively weak scaling of $A_\mathrm{eff}$ with $u_{nm}$, so the nonlinear coefficient increases for both larger $n$ and larger $m$, scaling approximately as $\gamma \propto u_{nm}^{\frac{9}{4}}$ for a fixed value of $n$, as shown in Fig.~\ref{fig:mode_properties}(b).

The balance of nonlinear and dispersive effects is encoded in the soliton order $N$, given by
\begin{equation}
    N = \sqrt{\frac{L_\mathrm{d}}{L_\mathrm{nl}}} = \sqrt{\frac{\gamma P_0 T_0^2}{\abs{\beta^{(2)}}}}\,,
\end{equation}
where $L_\mathrm{d} = T_0^2/\abs{\beta^{(2)}}$ and $L_\mathrm{nl} = (\gamma P_0)^{-1}$ are the dispersion and nonlinear lengths, respectively, with $P_0$ and $T_0$ the peak power and duration of the initial pulse, assuming the instantaneous power of the pulse follows the form $P(t) = P_0 \sech^2(t/T_0)$. A critical parameter in soliton self-compression and RDW emission in HCF is the length scale of the process. This can be approximated by the fission length, $L_\mathrm{f} = L_\mathrm{d}/N$ \cite{dudley_supercontinuum_2006}. Keeping $N$ and $\lambda_\mathrm{zd}$ constant, the increased anomalous dispersion in higher-order modes leads to a significant reduction in the fission length, as shown in Fig.~\ref{fig:mode_properties}(c). For $N$ to remain the same as the mode changes, the pulse energy has to be adjusted, because both the nonlinearity and the dispersion of the gas-filled waveguide change. As with the effective area, the energy corresponding to a certain soliton order increases for larger $n$ and decreases for larger $m$.

The propagation loss increases for both smaller cores and higher-order modes. However, this increase is less severe when increasing the mode order than when decreasing the core size, as $\alpha \propto u_{nm}^2/a^3$. The loss scales with $u_{nm}$ in the same way as the dispersion, so the ratio of the fission length to the loss length $L_\mathrm{loss} = 1/\alpha$ remains the same regardless of mode order (as long as the mode character is the same) but the required length of HCF is reduced. This suggests that propagation in higher-order modes makes soliton dynamics in HCF more readily achievable and in more compact systems than when using the fundamental mode.

\section{Experiment}
\begin{figure}
    \includegraphics[width=6in]{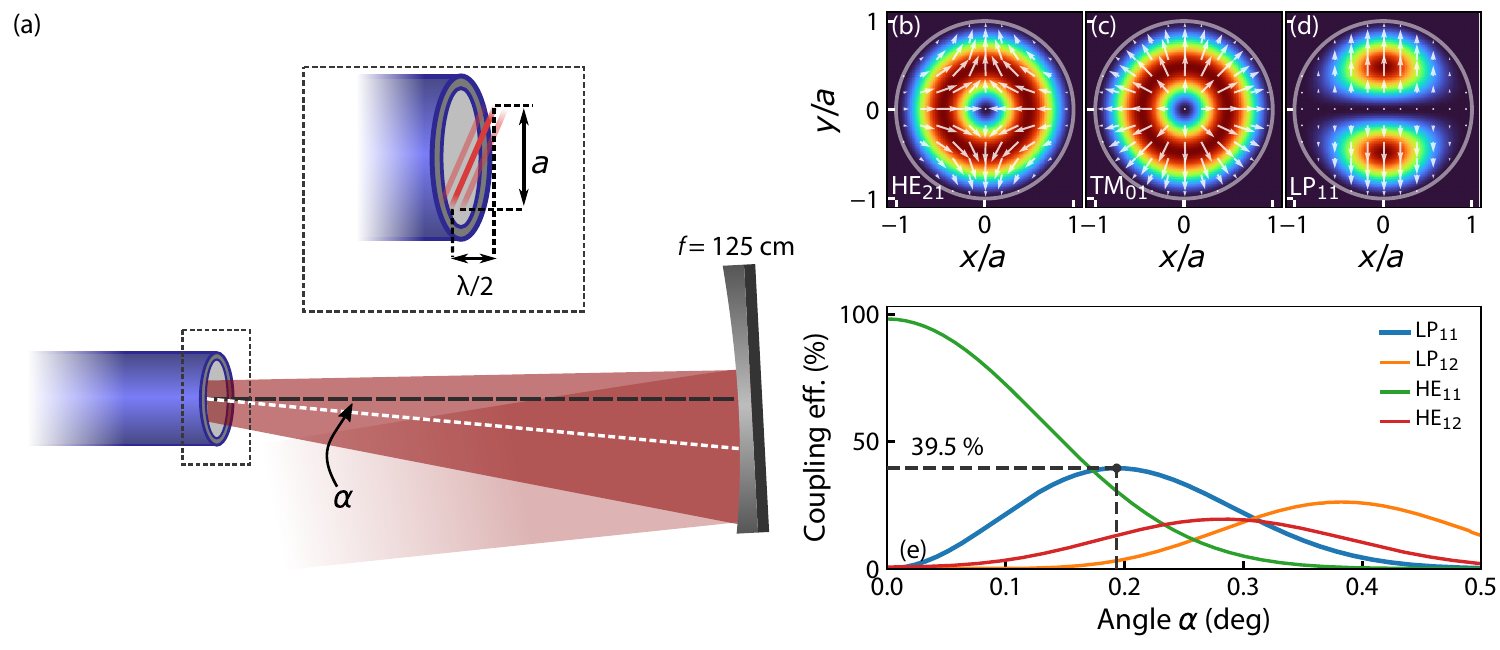}
    \caption{(a) Schematic representation of using oblique incidence to excite the LP$_{11}$ mode of an HCF. The inset shows how, for the correct angle of incidence, the path difference between the two halves of the core is $\lambda/2$, leading to a $\pi$ phase shift. (Note that the angle is exaggerated for visibility.) (b-d) Intensity profile (in false colour) and polarisation profile (white arrows) of the HE$_{21}$ and TM$_{01}$ modes, a superposition of which forms the LP$_{11}$ mode. The white circle shows the HCF core boundary. (e) Coupling efficiency into the most important double-lobe (LP$_{1m}$) and radially symmetric hybrid (HE$_{1m}$) modes as a function of tilt angle $\alpha$.}
    \label{fig:setup}
\end{figure}
The experimental setup is identical to that detailed in Ref.~\cite{travers_high-energy_2019}. In brief, a titanium-doped sapphire amplifier generates pulses at \SI{800}{\nm} central wavelength with a duration of \SI{30}{\fs}. These are spectrally broadened in a first HCF, which is filled with helium, and subsequently compressed to \SI{10}{\fs} duration using chirped mirrors and a wedge pair. After passing through a variable attenuator, the pulses are coupled into a second HCF with a core diameter of \SI{250}{\micro\meter} and a length of \SI{3}{\metre}, which is filled with argon. Both HCFs are stretched to eliminate bend loss~\cite{nagy_flexible_2008}.

Due to their complex field distribution, including radial phase variations and nodes, efficiently exciting the radially symmetric linearly polarised higher-order modes of an HCF (HE$_{1m}$ with $m>1$) with Gaussian-like pump beams requires complex beam shaping optics. As a simpler way to explore soliton dynamics in higher-order modes, we excite the linearly polarised double-lobe mode LP$_{11}$ instead. The two lobes of LP$_{11}$ have opposite phase, which can be approximately matched by coupling into the HCF with a tilted phase-front, as shown in Fig.~\ref{fig:setup}(a). The incidence angle for which this occurs can be approximated by $\alpha = \tan^{-1}\left(\lambda/2a\right)$, where $\lambda$ is the wavelength and $a$ is the core radius. For our parameters ($\lambda=\SI{800}{\nm}$, $a=\SI{125}{\micro\meter}$), this simple calculation results in $\alpha=\ang{0.183}$. A more accurate estimate of the optimal angle can be obtained by calculating the mode overlap integral,
\begin{equation}
    \label{eq:eta}
    \eta= \frac{
    \qty[\int_0^\infty\int_0^{2\pi}\bm{e}^*(r, \theta)\cdot \bm{E}(r, \theta)\,r\dd r\dd\theta]^2}
    {\int_0^\infty\int_0^{2\pi}\abs{\bm{e}(r,\theta)}^2\,r\dd r\dd\theta \int_0^\infty\int_0^{2\pi}\abs{\bm{E}(r,\theta)}^2\,r\dd r\dd\theta}\,,
\end{equation}
where $\bm{E}(r, \theta)$ is the field distribution of the incoming laser beam. The LP$_{11}$ mode can be expressed as a superposition of HE$_{21}$ and TM$_{01}$ [see Fig.~\ref{fig:setup}(b-d)], which have the same value of $u_{nm}$ ($\sim 3.831$) and hence identical dispersion. A Gaussian beam with linear polarisation along the $y$ axis which is incident at an angle $\alpha$ to the $z$ axis in the $y$--$z$ plane can be described as
\begin{equation}
    \bm{E}(r, \theta) = E_0 \hat{\bm{y}} \exp{-\frac{r^2}{w_0^2}\left(\cos^2{\theta}+\frac{\sin^2{\theta}}{\cos^2{\alpha}}\right) + i rk_0\sin{\alpha}\sin{\theta}}\,,
\end{equation}
where $w_0$ is the $1/\mathrm{e}^2$ radius of the beam and $\hat{\bm{y}}$ is the unit vector in the $y$-direction. As shown in Fig.~\ref{fig:setup}(e), for a Gaussian laser beam with a radius of $w_0=0.64a$ at \SI{800}{\nm} (which maximises coupling to HE$_{11}$ for normal incidence), an incidence of angle of \ang{0.193} maximises the overlap with LP$_{11}$, in good agreement with the simple calculation. At this angle, \SI{39.5}{\percent} of the incident energy is coupled into the LP$_{11}$ mode. For our focusing geometry, which consists of a single $f=\SI{1.25}{\metre}$ concave spherical mirror, a translation of the beam on the focusing optic of \SI{4.2}{\mm} is required. While simple, this configuration does not lead to coupling purely into LP$_{11}$---approximately \SI{34}{\percent} of the energy is coupled into the fundamental HE$_{11}$ mode, \SI{12}{\percent} into HE$_{12}$, \SI{3}{\percent} into LP$_{12}$, and the remainder into other high-order modes of HE and TM character.

\begin{figure}
    \centering
    \includegraphics[width=6in]{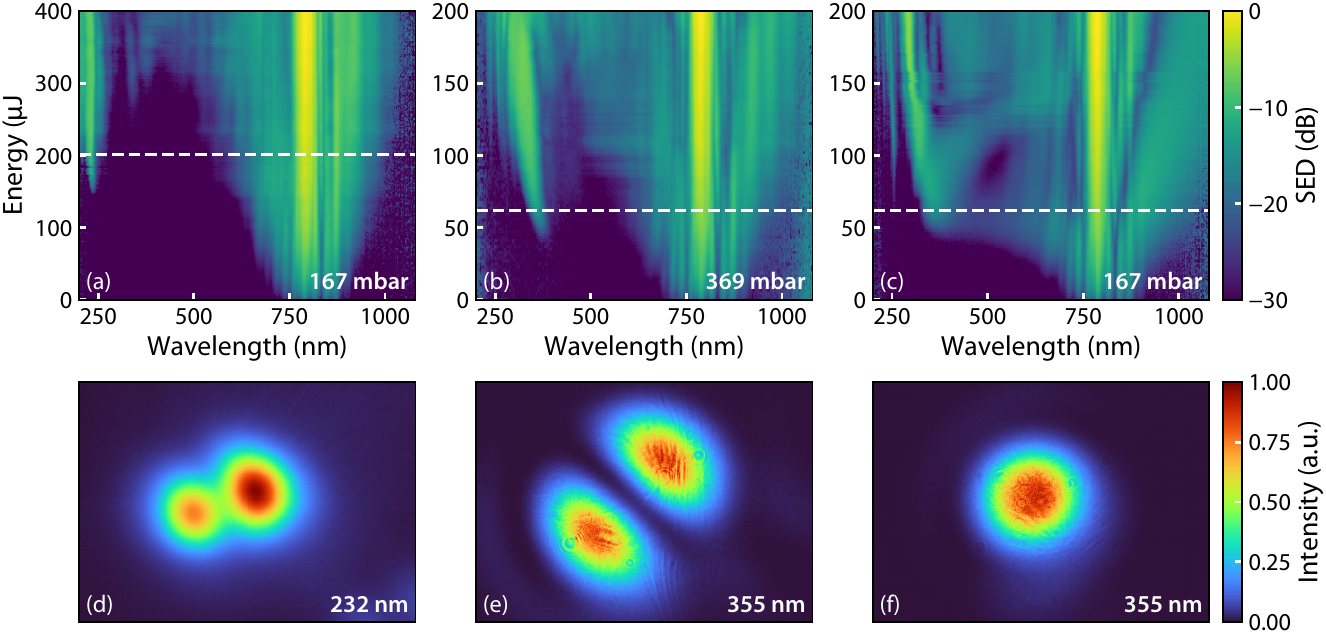}
    \caption{Experimental demonstration of self-compression and RDW emission in a higher-order mode. Top row: output spectrum of the HCF (on a logarithmic colour scale) as a function of incident pump energy at the indicated argon pressures when coupling at an angle (a, b) and when coupling along the optical axis (c). Bottom row: far-field beam profile of the dispersive wave, bandpass-filtered around the indicated wavelengths, when pumping with the energies indicated by the white dashed line in the top row.}
    \label{fig:exp_escan}
\end{figure}

Figure \ref{fig:exp_escan} shows experimentally measured output spectra of the second HCF as a function of incident pump energy for three combinations of argon pressure and incidence angle. In all three cases, the evolution of the output spectrum shows typical features of soliton self-compression and RDW emission: rapid spectral broadening around the pump wavelength at low energy followed by the emergence and subsequent blue-shift of the RDW peak at the phase-matched wavelength, in this case in the UV. However, the spectral broadening around the pump wavelength is less pronounced when pumping in the LP$_{11}$ mode [Fig.~\ref{fig:exp_escan}(a, b)] than when pumping in the fundamental mode [Fig.~\ref{fig:exp_escan}(c)], even at energies where RDW emission occurs. Furthermore, the RDW peak appears at a different wavelength for the same gas pressure [compare Fig.~\ref{fig:exp_escan}(a) and (c)]. The bright, narrowband feature around the pump wavelength of \SI{800}{\nm}, which does not change appreciably as a function of energy, is the result of poor pulse contrast on the initial pulses generated by our laser system~\cite{brahms_high-energy_2019}.

The bottom row in Fig.~\ref{fig:exp_escan} shows far-field beam profiles of the dispersive-wave pulse, measured at an energy near the initial emergence of the RDW [white dashed lines in Fig.~\ref{fig:exp_escan}(a-c)] after bandpass filtering with \SI{10}{\nm} bandwidth around the spectral peak. The dispersive waves pumped in the higher-order mode clearly show the characteristic two-lobe pattern of the LP$_{11}$ mode. The beam at \SI{232}{\nm} [Fig.~\ref{fig:exp_escan}(d)] appears blurred; this is likely the result of cross-talk between camera pixels caused by the high intensity on the sensor, which is required because the quantum efficiency of our camera is very low in this wavelength region. Small-scale features in Fig.~\ref{fig:exp_escan}(e) and (f) are caused by imperfections of the bandpass filter.

\begin{figure}
    \centering
    \includegraphics[width=6in]{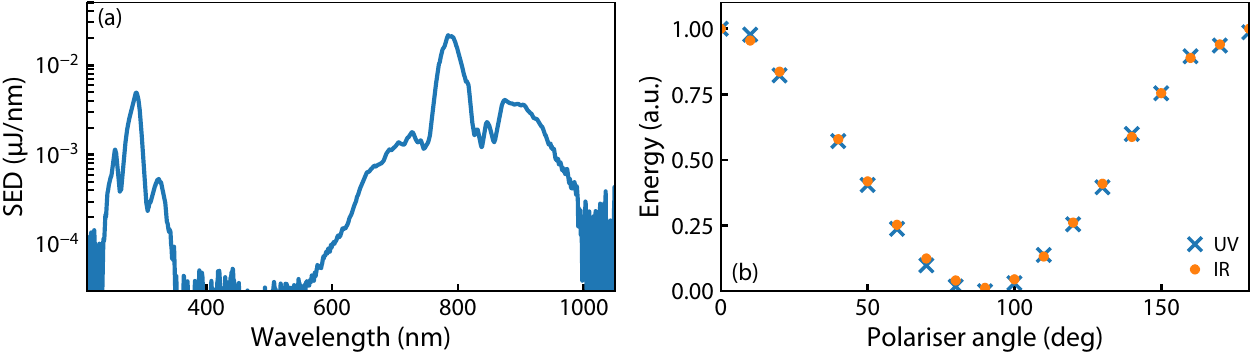}
    \caption{(a) Output spectrum at an argon pressure of \SI{263}{\milli\bar} and an incident pump energy of \SI{236}{\micro\joule}. (b) Energy (on a normalised scale) in the UV ($<\SI{350}{\nm}$) and IR ($>\SI{550}{\nm}$) parts of the spectrum as a function of polariser rotation angle, with \ang{0} corresponding to vertical orientation of the polariser.}
    \label{fig:pol}
\end{figure}

The RDW beam profiles in Fig.~\ref{fig:exp_escan}(d) and (e) contain no noticeable contribution from other modes, including the fundamental mode, even though our focusing geometry excites a variety of modes at the HCF entrance. This is a consequence of the mode-selection property of RDW emission: since the phase-matching to dispersive waves depends critically on the modal dispersion contribution, energy is efficiently transferred to the spectral band of the dispersive wave only in the phase-matched mode. Another consequence of this is the near-perfect beam quality of fundamental-mode dispersive-waves pulses, as previously observed~\cite{travers_high-energy_2019} and also shown in Fig.~\ref{fig:exp_escan}(f). The lobes of the dispersive-wave modes shown in Fig.~\ref{fig:exp_escan}(d) and (e) are tilted with respect to the example shown in Fig.~\ref{fig:setup}(d), even though we translate the beam purely vertically on the focusing mirror to excite the LP$_{11}$ mode. We attribute this to aberrations in our incident laser beam---such as astigmatism caused by off-axis reflection from a spherical mirror---which may change the coupling conditions. We find that small adjustments to the focusing mirror can rotate the mode of the dispersive wave but simultaneously affect the coupling efficiency and RDW emission. The mode images shown in Fig.~\ref{fig:exp_escan} represent the coupling conditions which maximise the RDW energy at the HCF output. Figure~\ref{fig:pol} shows the polarisation direction of the pulse as characterised by measuring the output spectrum after a broadband Rochon prism polariser as a function of polariser angle. Although the mode profile is rotated with respect to the calculated mode, the polarisation is still purely linear and aligned with that of the incident beam.

\begin{figure}
    \centering
    \includegraphics[width=6in]{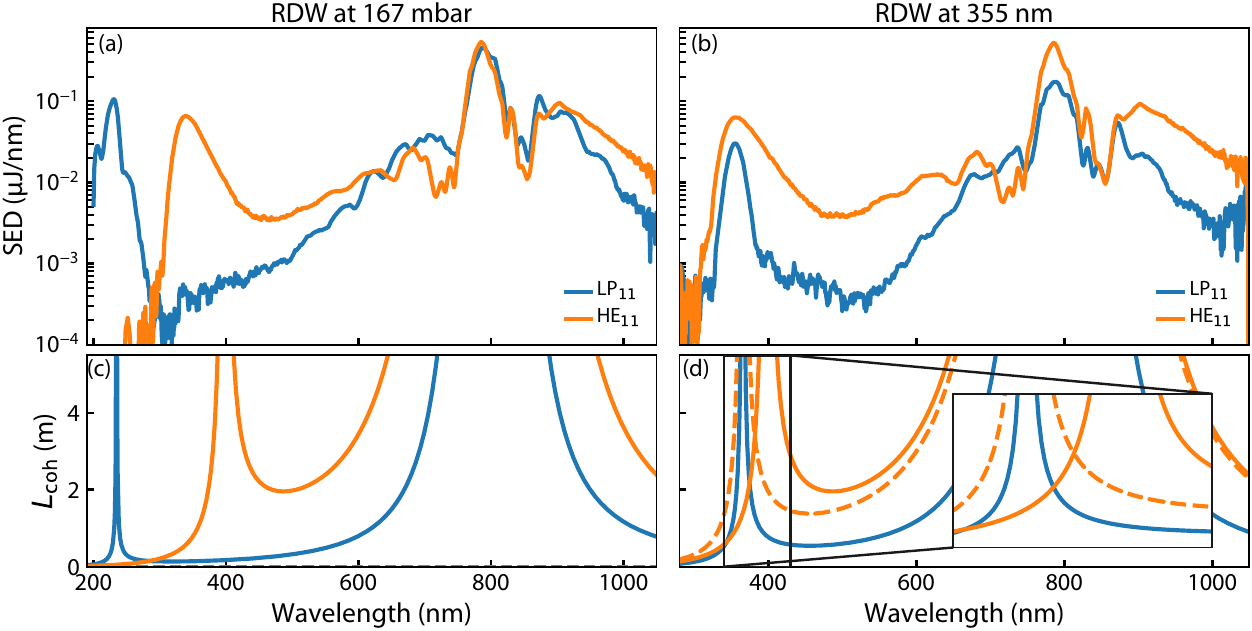}
    \caption{(a) Output spectra at the same pressure of \SI{167}{\milli\bar} and pump energy of \SI{202}{\micro\joule} and \SI{62}{\micro\joule} for the LP$_{11}$ and HE$_{11}$ mode, respectively, corresponding to the beam profiles shown in Fig.~\ref{fig:exp_escan}(d) and (f). (b) Comparison between output spectra for RDW emission at \SI{355}{\nm} in the two modes. The blue line shows the LP$_{11}$ spectrum at \SI{369}{\milli\bar} argon pressure and \SI{67}{\micro\joule} pump energy and the orange line shows the HE$_{11}$ spectrum at \SI{167}{\milli\bar} pressure and \SI{70}{\micro\joule} of pump energy. (c) Calculated coherence length for RDW emission in the two modes at \SI{167}{\milli\bar}. (d) Calculated coherence length for RDW emission at the two different pressures shown in (b).}
    \label{fig:exp_lines}
\end{figure}

In Figure \ref{fig:exp_lines}, we investigate the differences between RDW emission in fundamental and higher-order modes in more detail. Fig.~\ref{fig:exp_lines}(a) shows spectra at \SI{167}{\milli\bar}, corresponding to the beam profiles shown in Fig.~\ref{fig:exp_escan}(d) and (f). In addition to the clear difference in RDW emission wavelength (\SI{232}{\nm} in LP$_{11}$ as opposed to \SI{340}{\nm} in HE$_{11}$), the reduced coupling efficiency and stronger propagation loss in the higher-order mode is evident in the fact that the spectrum near the pump is generally similar, despite the LP$_{11}$ RDW being pumped with three times the energy (\SI{202}{\micro\joule} as compared to \SI{67}{\micro\joule}). Fig.~\ref{fig:exp_lines}(b) compares RDW emission for the parameters which place the RDW peak at \SI{355}{\nm} in both modes. To match the phase-matching point, the pressure is increased to \SI{369}{\milli\bar} for pumping in LP$_{11}$. One obvious difference is in the overall spectral energy density, which is significantly lower in LP$_{11}$. However, it is important to note that the total input energy required to obtain RDW emission is similar (\SI{67}{\micro\joule} and \SI{70}{\micro\joule} in LP$_{11}$ and HE$_{11}$, respectively), despite stronger losses and lower coupling efficiency. This is because, in addition to the generally stronger nonlinearity for higher-order modes [see Fig.~\ref{fig:mode_properties}(b)], the effective area of the LP$_{11}$ mode is a factor of $2/3$ smaller than that of HE$_{21}$ and TM$_{01}$, further increasing the nonlinearity.

Other differences appear in the spectral shape of the continuum and the RDW peak. In the higher-order mode, the dip in the spectrum between pump and RDW wavelengths is more pronounced, leaving the RDW peak more isolated. Additionally, the bandwidth of the RDW is significantly smaller in the higher-order mode: the full width at half-maximum (FWHM) is \SI{24}{\nm} as compared to \SI{52}{\nm} in HE$_{11}$. Both of these observations, as well as the spectral blue-shift of the higher-order-mode RDW for a fixed pressure, can be attributed to phase-matching effects. RDW phase-matching can be described approximately by treating the self-compressing pulse as a soliton experiencing no dispersion and writing the propagation constants of the soliton, $\beta_\mathrm{s}$, and that of the dispersive wave, $\beta_\mathrm{dw}$, as
\begin{align}
    \beta_\mathrm{s}(\omega) &= \beta^{(0)} + \beta^{(1)} \Delta\omega  \label{eq:bsol}\\ 
    \beta_\mathrm{dw}(\omega) &= \beta^{(0)} + \beta^{(1)} \Delta\omega + \frac{\beta^{(2)}}{2} \Delta\omega^2 + \frac{\beta^{(3)}}{6} \Delta\omega^3 + \cdots\,,
    \label{eq:blin}
\end{align}
where $\beta^{(i)} = \partial_\omega^i\beta(\omega)\vert_{\omega_0}$ and $\Delta\omega = \omega-\omega_0$ is the frequency detuning from the central frequency of the soliton $\omega_0$. The phase mismatch is then $\Delta\beta(\omega) = \beta_\mathrm{dw}(\omega) - \beta_\mathrm{s}(\omega)$. Figure \ref{fig:exp_lines}(c) and (d) show the coherence length $L_\mathrm{coh} = \pi/\Delta\beta(\omega)$ in the LP$_{11}$ and HE$_{11}$ modes for the conditions in Fig.~\ref{fig:exp_lines}(a) and (b), respectively. The most obvious difference in Fig.~\ref{fig:exp_lines}(c) is that the coherence length diverges at different UV wavelengths, indicating that the phase-matching point for RDW emission has shifted. This is due to the stronger anomalous dispersion, in a similar manner to the dispersive-wave blue-shift for lower pressures \cite{mak_tunable_2013,travers_high-energy_2019}.

Away from the pump wavelength and the RDW phase-matching points, the coherence length between soliton and dispersive wave is far shorter in the higher-order mode. As a consequence, these frequency components are less efficiently generated, leading to the larger dip in the supercontinuum spectrum as observed in the experiment. The inset in Fig.~\ref{fig:exp_lines}(d) shows a detailed view around the phase-matching point in the UV. The key feature is the difference in bandwidth of the two peaks. For the fundamental mode, the phase mismatch changes more slowly around the phase-matching point, leading to a larger phase-matching bandwidth. This explains the difference in observed RDW bandwidth in Fig.~\ref{fig:exp_lines}(b). The small discrepancy between phase-matching wavelengths, which is not observed in the experiment for these pressures, can be attributed to the fact that eq.~\ref{eq:bsol} neglects the nonlinear contribution to $\beta_\mathrm{s}$. This is difficult to estimate for the mixed-mode input in our experiment but would certainly differ between the two modes. The dashed line in Fig.~\ref{fig:exp_lines}(d) shows the coherence length for the pressure at which the phase-matched wavelengths according to eqs.~\ref{eq:bsol} and \ref{eq:blin} coincide; the difference in overall coherence length and phase-matching bandwidth is present here as well.

\section{Numerical simulations}
To investigate the nonlinear dynamics in more detail and confirm the predictions of the scaling rules laid out in section \ref{sec:properties}, we numerically simulate self-compression and RDW emission in our system. The numerical model and its implementation are described in detail elsewhere \cite{brahms_lunajl_2021,travers_high-energy_2019}. We model the fully space- and polarisation-resolved multimode propagation including the modal dispersion and loss, the Kerr effect, and photoionisation and plasma dynamics.

\begin{figure}
    \centering
    \includegraphics[width=6in]{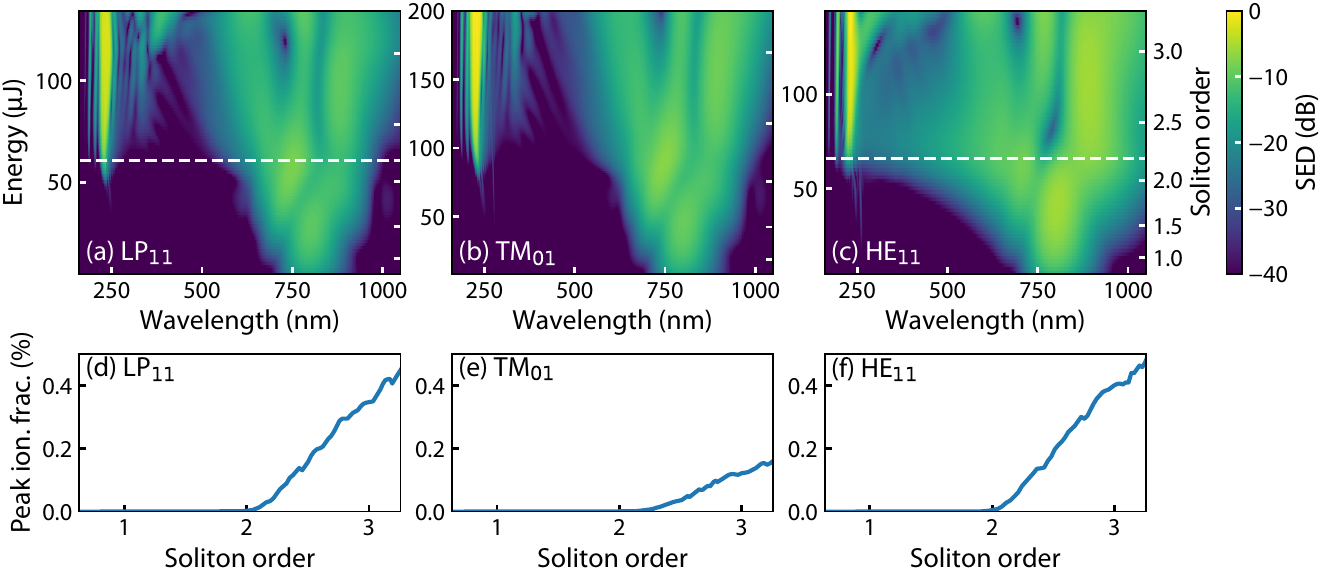}
    \caption{(a-c) Simulated output spectra as a function of coupled pump energy when pumping in the modes LP$_{11}$ (a), TM$_{01}$ (b), and HE$_{11}$ (c) for a fixed zero-dispersion wavelength of $\lambda_\mathrm{zd}=\SI{500}{\nm}$, corresponding to \SI{167}{\milli\bar} argon pressure in LP$_{11}$ and TM$_{01}$ and \SI{67}{\milli\bar} in HE$_{11}$. The right axis in (c) is common to all top-row plots and shows the soliton order. The white dashed lines in (a) and (c) indicate the energies for which the time-domain evolution is shown in Fig.~\ref{fig:sim_singles}. (d-f) maximum ionisation fraction reached during propagation for the same range of soliton orders shown in (a-c).}
    \label{fig:sim_differentModes}
\end{figure}

Figure \ref{fig:sim_differentModes} shows simulated energy-dependent output spectra of the gas-filled HCF for a fixed zero-dispersion wavelength of \SI{500}{\nm} in three different modes: in addition to the two modes observed experimentally [Fig.~\ref{fig:sim_differentModes}(a) and (c)], we simulate the propagation dynamics in the pure TM$_{01}$ mode [Fig.~\ref{fig:sim_differentModes}(b)]. We consider an ideal Gaussian laser pulse with a FWHM duration of \SI{10}{\femto\second}. For propagation in LP$_{11}$, we place half of the energy into HE$_{21}$ and TM$_{01}$, respectively, and include higher-modes up to HE$_{24}$ and TM$_{04}$. For propagation in the pure modes, we include higher-order modes up to TM$_{04}$ and HE$_{14}$, respectively.

The results shown in Fig.~\ref{fig:sim_differentModes}(a) and (c) reproduce the features observed experimentally, including the more pronounced dip in the supercontinuum between the pump and RDW peaks for propagation in LP$_{11}$. Since we adjust the pressure to fix the zero-dispersion wavelength, we obtain RDW emission in the same spectral region. Similarly to the experimental results for RDW emission at \SI{355}{\nm}, the energy required to first obtain RDW emission is closely matched---\SI{60}{\micro\joule} and \SI{66}{\micro\joule} in LP$_{11}$ and HE$_{11}$, respectively---because the small effective area of LP$_{11}$ counteracts the general increase in required energy when moving to higher-order modes: the effective area of the LP$_{11}$ mode with a core radius of \SI{125}{\micro\meter} is $\sim\SI{0.0234}{\square\mm}$, as compared to $\sim\SI{0.0211}{\square\mm}$ for HE$_{11}$. In combination with the matched dispersion, this close correspondence accounts for the similarity in energy scale.

The similar peak intensity in LP$_{11}$ and HE$_{11}$ is also reflected in the peak ionisation fraction reached during propagation, as shown in Fig.~\ref{fig:sim_differentModes}(d) and (f), which is closely matched between the two cases. Propagation in TM$_{01}$ presents a different picture, however: although the overall dynamics are very similar [compare Fig.~\ref{fig:sim_differentModes}(a) and (b)], the electron density is significantly lower when self-compression and RDW emission occur in TM$_{01}$, as shown in Fig.~\ref{fig:sim_differentModes}(e). This is despite the higher energy required for the same dynamics in this mode, as expected from the factor of $3/2$ in the effective area as compared to LP$_{11}$, with RDW emission first appearing around \SI{90}{\micro\joule}. Modes of higher azimuthal order thus allow for more energetic self-compression while suppressing photoionisation, which may be advantageous for the generation of very high-energy sub-cycle transients and resonant dispersive waves.

\begin{figure}
    \centering
    \includegraphics[width=4.5in]{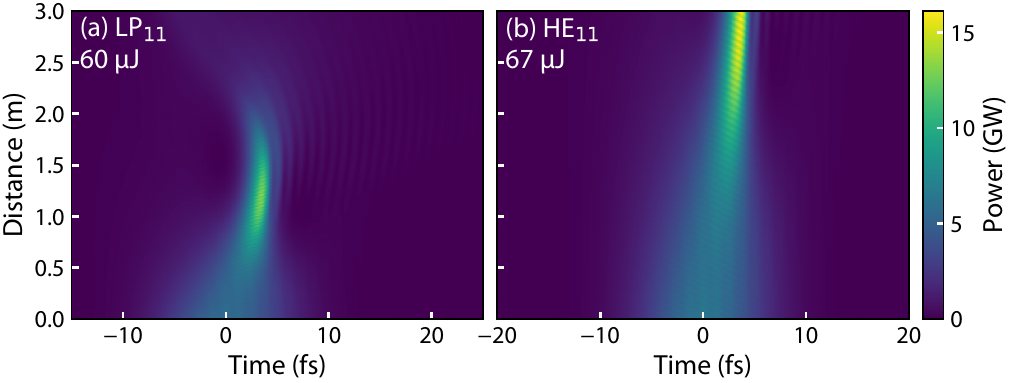}
    \caption{Time-domain evolution of the self-compressing pulse for a soliton order of $N=2.2$ in the modes LP$_{11}$ (a) and HE$_{11}$ (b) for a fixed zero-dispersion wavelength of \SI{500}{\nm}.}
    \label{fig:sim_singles}
\end{figure}

As shown in Fig.~\ref{fig:mode_properties}(c), stronger dispersion in higher-order modes should lead to faster self-compression for a given soliton order. This prediction is confirmed in Fig.~\ref{fig:sim_singles}, which shows the evolution of the pulse shape (in a reference frame moving with the group velocity at the pump) during propagation in LP$_{11}$ and HE$_{11}$ for the energy at which RDW emission first occurs, corresponding to a soliton order of $N=2.2$. When propagating in the higher-order mode, the pulse clearly self-compresses much more rapidly, reaching its maximal peak power within less than \SI{1.5}{\meter}, as opposed to near the end of the waveguide. The peak power at the self-compression point is \SI{13.3}{\giga\watt} in LP$_{11}$, whereas in HE$_{11}$ it reaches \SI{16.1}{\giga\watt}. From the ratio of pump energies, a maximal peak power of \SI{14.4}{\giga\watt} would be expected in LP$_{11}$. The discrepancy is not due to a difference in the self-compression process: the two pulses self-compress to near-identical duration, \SI{1.76}{\fs} for HE$_{11}$ and \SI{1.77}{\fs} for LP$_{11}$. Instead, a small difference in the propagation loss causes the pulse to lose more energy in the higher-order mode. This difference stems from the fact that the LP$_{11}$ mode is constituted from two modes of different character, HE$_{21}$ and TM$_{01}$. HE$_{21}$ has the same HE character as the fundamental mode, so the loss scaling detailed in section \ref{sec:properties} applies. TM$_{01}$, however, is more lossy due to a difference in the $\nu_n(\omega)$ term even though the value of $u_{nm}$ is identical \cite{marcatili_hollow_1964}. Using the average propagation loss of the two modes with the scaled fission length predicts a maximum peak power of \SI{13.5}{\giga\watt} for LP$_{11}$, in good agreement with the simulation results. It should be noted that due to this difference in loss, LP$_{11}$ is not strictly speaking a mode of the waveguide, as even an input field with purely this field distribution will not remain pure upon propgation.

\section{Conclusions}
In summary, we have explored the properties of soliton self-compression in the higher-order modes of gas-filled hollow capillary fibres using simple scaling rules, experimental investigations, and numerical simulations. Our first key finding is that the use of higher-order modes can shorten the required waveguide without incurring a loss penalty, in contrast to a reduction in core size \cite{brahms_high-energy_2019}. Secondly, resonant dispersive waves generated in higher-order modes show exceptional mode purity, even in the presence of pulses in multiple modes interacting nonlinearly. Thirdly, the stronger anomalous dispersion contribution from higher-order modes affects the central wavelength and bandwidth of dispersive waves as well the shape of the supercontinuum between the dispersive wave and the pump. Finally, self-compression in higher-order modes can be achieved at significantly higher energy while simultaneously suppressing photoionisation effects as compared to equivalent parameters in the fundamental mode. We expect that our results will be useful for the design of more flexible, powerful, and compact systems to generate sub-cycle field transients as well as widely tuneable few-femtosecond pulses.

\section*{Acknowledgements}
This work was funded by the European Research Council under the European Union’s Horizon 2020 Research and Innovation program: Starting Grant agreement HISOL, No.~679649; Proof of Concept Grant agreement ULIGHT, No.~899900; and by the United Kingdom's Engineering and Physical Sciences Research Council: Grant agreement EP/T020903/1

\bibliography{bibliography}

\begin{thebibliography}{17}%
\makeatletter
\providecommand \@ifxundefined [1]{%
 \@ifx{#1\undefined}
}%
\providecommand \@ifnum [1]{%
 \ifnum #1\expandafter \@firstoftwo
 \else \expandafter \@secondoftwo
 \fi
}%
\providecommand \@ifx [1]{%
 \ifx #1\expandafter \@firstoftwo
 \else \expandafter \@secondoftwo
 \fi
}%
\providecommand \natexlab [1]{#1}%
\providecommand \enquote  [1]{``#1''}%
\providecommand \bibnamefont  [1]{#1}%
\providecommand \bibfnamefont [1]{#1}%
\providecommand \citenamefont [1]{#1}%
\providecommand \href@noop [0]{\@secondoftwo}%
\providecommand \href [0]{\begingroup \@sanitize@url \@href}%
\providecommand \@href[1]{\@@startlink{#1}\@@href}%
\providecommand \@@href[1]{\endgroup#1\@@endlink}%
\providecommand \@sanitize@url [0]{\catcode `\\12\catcode `\$12\catcode
  `\&12\catcode `\#12\catcode `\^12\catcode `\_12\catcode `\%12\relax}%
\providecommand \@@startlink[1]{}%
\providecommand \@@endlink[0]{}%
\providecommand \url  [0]{\begingroup\@sanitize@url \@url }%
\providecommand \@url [1]{\endgroup\@href {#1}{\urlprefix }}%
\providecommand \urlprefix  [0]{URL }%
\providecommand \Eprint [0]{\href }%
\providecommand \doibase [0]{https://doi.org/}%
\providecommand \selectlanguage [0]{\@gobble}%
\providecommand \bibinfo  [0]{\@secondoftwo}%
\providecommand \bibfield  [0]{\@secondoftwo}%
\providecommand \translation [1]{[#1]}%
\providecommand \BibitemOpen [0]{}%
\providecommand \bibitemStop [0]{}%
\providecommand \bibitemNoStop [0]{.\EOS\space}%
\providecommand \EOS [0]{\spacefactor3000\relax}%
\providecommand \BibitemShut  [1]{\csname bibitem#1\endcsname}%
\let\auto@bib@innerbib\@empty
\bibitem [{\citenamefont {Travers}\ \emph {et~al.}(2011)\citenamefont
  {Travers}, \citenamefont {Chang}, \citenamefont {Nold}, \citenamefont
  {Joly},\ and\ \citenamefont {St. J.~Russell}}]{travers_ultrafast_2011}%
  \BibitemOpen
  \bibfield  {author} {\bibinfo {author} {\bibfnamefont {J.~C.}\ \bibnamefont
  {Travers}}, \bibinfo {author} {\bibfnamefont {W.}~\bibnamefont {Chang}},
  \bibinfo {author} {\bibfnamefont {J.}~\bibnamefont {Nold}}, \bibinfo {author}
  {\bibfnamefont {N.~Y.}\ \bibnamefont {Joly}},\ and\ \bibinfo {author}
  {\bibfnamefont {P.}~\bibnamefont {St. J.~Russell}},\ }\bibfield  {title}
  {\bibinfo {title} {Ultrafast nonlinear optics in gas-filled hollow-core
  photonic crystal fibers [{{Invited}}]},\ }\href
  {https://doi.org/10.1364/JOSAB.28.000A11} {\bibfield  {journal} {\bibinfo
  {journal} {Journal of the Optical Society of America B}\ }\textbf {\bibinfo
  {volume} {28}},\ \bibinfo {pages} {A11} (\bibinfo {year} {2011})}\BibitemShut
  {NoStop}%
\bibitem [{\citenamefont {Russell}\ \emph {et~al.}(2014)\citenamefont
  {Russell}, \citenamefont {H{\"o}lzer}, \citenamefont {Chang}, \citenamefont
  {Abdolvand},\ and\ \citenamefont {Travers}}]{russell_hollow-core_2014}%
  \BibitemOpen
  \bibfield  {author} {\bibinfo {author} {\bibfnamefont {P.~S.~J.}\
  \bibnamefont {Russell}}, \bibinfo {author} {\bibfnamefont {P.}~\bibnamefont
  {H{\"o}lzer}}, \bibinfo {author} {\bibfnamefont {W.}~\bibnamefont {Chang}},
  \bibinfo {author} {\bibfnamefont {A.}~\bibnamefont {Abdolvand}},\ and\
  \bibinfo {author} {\bibfnamefont {J.~C.}\ \bibnamefont {Travers}},\
  }\bibfield  {title} {\bibinfo {title} {Hollow-core photonic crystal fibres
  for gas-based nonlinear optics},\ }\href
  {https://doi.org/10.1038/nphoton.2013.312} {\bibfield  {journal} {\bibinfo
  {journal} {Nature Photonics}\ }\textbf {\bibinfo {volume} {8}},\ \bibinfo
  {pages} {278} (\bibinfo {year} {2014})}\BibitemShut {NoStop}%
\bibitem [{\citenamefont {Nagy}\ \emph {et~al.}(2021)\citenamefont {Nagy},
  \citenamefont {Simon},\ and\ \citenamefont {Veisz}}]{nagy_high-energy_2021}%
  \BibitemOpen
  \bibfield  {author} {\bibinfo {author} {\bibfnamefont {T.}~\bibnamefont
  {Nagy}}, \bibinfo {author} {\bibfnamefont {P.}~\bibnamefont {Simon}},\ and\
  \bibinfo {author} {\bibfnamefont {L.}~\bibnamefont {Veisz}},\ }\bibfield
  {title} {\bibinfo {title} {High-energy few-cycle pulses: Post-compression
  techniques},\ }\href {https://doi.org/10.1080/23746149.2020.1845795}
  {\bibfield  {journal} {\bibinfo  {journal} {Advances in Physics: X}\ }\textbf
  {\bibinfo {volume} {6}},\ \bibinfo {pages} {1845795} (\bibinfo {year}
  {2021})}\BibitemShut {NoStop}%
\bibitem [{\citenamefont {Joly}\ \emph {et~al.}(2011)\citenamefont {Joly},
  \citenamefont {Nold}, \citenamefont {Chang}, \citenamefont {H{\"o}lzer},
  \citenamefont {Nazarkin}, \citenamefont {Wong}, \citenamefont {Biancalana},\
  and\ \citenamefont {Russell}}]{joly_bright_2011}%
  \BibitemOpen
  \bibfield  {author} {\bibinfo {author} {\bibfnamefont {N.~Y.}\ \bibnamefont
  {Joly}}, \bibinfo {author} {\bibfnamefont {J.}~\bibnamefont {Nold}}, \bibinfo
  {author} {\bibfnamefont {W.}~\bibnamefont {Chang}}, \bibinfo {author}
  {\bibfnamefont {P.}~\bibnamefont {H{\"o}lzer}}, \bibinfo {author}
  {\bibfnamefont {A.}~\bibnamefont {Nazarkin}}, \bibinfo {author}
  {\bibfnamefont {G.~K.~L.}\ \bibnamefont {Wong}}, \bibinfo {author}
  {\bibfnamefont {F.}~\bibnamefont {Biancalana}},\ and\ \bibinfo {author}
  {\bibfnamefont {P.~S.~J.}\ \bibnamefont {Russell}},\ }\bibfield  {title}
  {\bibinfo {title} {Bright spatially coherent wavelength-tunable deep-{{UV}}
  laser source using an {{Ar}}-filled photonic crystal fiber},\ }\href
  {https://doi.org/10.1103/PhysRevLett.106.203901} {\bibfield  {journal}
  {\bibinfo  {journal} {Physical Review Letters}\ }\textbf {\bibinfo {volume}
  {106}},\ \bibinfo {pages} {203901} (\bibinfo {year} {2011})}\BibitemShut
  {NoStop}%
\bibitem [{\citenamefont {Mak}\ \emph {et~al.}(2013)\citenamefont {Mak},
  \citenamefont {Travers}, \citenamefont {H{\"o}lzer}, \citenamefont {Joly},\
  and\ \citenamefont {Russell}}]{mak_tunable_2013}%
  \BibitemOpen
  \bibfield  {author} {\bibinfo {author} {\bibfnamefont {K.~F.}\ \bibnamefont
  {Mak}}, \bibinfo {author} {\bibfnamefont {J.~C.}\ \bibnamefont {Travers}},
  \bibinfo {author} {\bibfnamefont {P.}~\bibnamefont {H{\"o}lzer}}, \bibinfo
  {author} {\bibfnamefont {N.~Y.}\ \bibnamefont {Joly}},\ and\ \bibinfo
  {author} {\bibfnamefont {P.~S.~J.}\ \bibnamefont {Russell}},\ }\bibfield
  {title} {\bibinfo {title} {Tunable vacuum-{{UV}} to visible ultrafast pulse
  source based on gas-filled {{Kagome}}-{{PCF}}.},\ }\href
  {https://doi.org/10.1364/OE.21.010942} {\bibfield  {journal} {\bibinfo
  {journal} {Optics Express}\ }\textbf {\bibinfo {volume} {21}},\ \bibinfo
  {pages} {10942} (\bibinfo {year} {2013})}\BibitemShut {NoStop}%
\bibitem [{\citenamefont {Ermolov}\ \emph {et~al.}(2015)\citenamefont
  {Ermolov}, \citenamefont {Mak}, \citenamefont {Frosz}, \citenamefont
  {Travers},\ and\ \citenamefont {Russell}}]{ermolov_supercontinuum_2015}%
  \BibitemOpen
  \bibfield  {author} {\bibinfo {author} {\bibfnamefont {A.}~\bibnamefont
  {Ermolov}}, \bibinfo {author} {\bibfnamefont {K.~F.}\ \bibnamefont {Mak}},
  \bibinfo {author} {\bibfnamefont {M.~H.}\ \bibnamefont {Frosz}}, \bibinfo
  {author} {\bibfnamefont {J.~C.}\ \bibnamefont {Travers}},\ and\ \bibinfo
  {author} {\bibfnamefont {P.~S.~J.}\ \bibnamefont {Russell}},\ }\bibfield
  {title} {\bibinfo {title} {Supercontinuum generation in the vacuum
  ultraviolet through dispersive-wave and soliton-plasma interaction in a
  noble-gas-filled hollow-core photonic crystal fiber},\ }\href
  {https://doi.org/10.1103/PhysRevA.92.033821} {\bibfield  {journal} {\bibinfo
  {journal} {Physical Review A}\ }\textbf {\bibinfo {volume} {92}},\ \bibinfo
  {pages} {033821} (\bibinfo {year} {2015})}\BibitemShut {NoStop}%
\bibitem [{\citenamefont {Travers}\ \emph {et~al.}(2019)\citenamefont
  {Travers}, \citenamefont {Grigorova}, \citenamefont {Brahms},\ and\
  \citenamefont {Belli}}]{travers_high-energy_2019}%
  \BibitemOpen
  \bibfield  {author} {\bibinfo {author} {\bibfnamefont {J.~C.}\ \bibnamefont
  {Travers}}, \bibinfo {author} {\bibfnamefont {T.~F.}\ \bibnamefont
  {Grigorova}}, \bibinfo {author} {\bibfnamefont {C.}~\bibnamefont {Brahms}},\
  and\ \bibinfo {author} {\bibfnamefont {F.}~\bibnamefont {Belli}},\ }\bibfield
   {title} {\bibinfo {title} {High-energy pulse self-compression and
  ultraviolet generation through soliton dynamics in hollow capillary fibres},\
  }\href {https://doi.org/10.1038/s41566-019-0416-4} {\bibfield  {journal}
  {\bibinfo  {journal} {Nature Photonics}\ }\textbf {\bibinfo {volume} {13}},\
  \bibinfo {pages} {547} (\bibinfo {year} {2019})}\BibitemShut {NoStop}%
\bibitem [{\citenamefont {Brahms}\ \emph {et~al.}(2020)\citenamefont {Brahms},
  \citenamefont {Belli},\ and\ \citenamefont {Travers}}]{brahms_infrared_2020}%
  \BibitemOpen
  \bibfield  {author} {\bibinfo {author} {\bibfnamefont {C.}~\bibnamefont
  {Brahms}}, \bibinfo {author} {\bibfnamefont {F.}~\bibnamefont {Belli}},\ and\
  \bibinfo {author} {\bibfnamefont {J.~C.}\ \bibnamefont {Travers}},\
  }\bibfield  {title} {\bibinfo {title} {Infrared attosecond field transients
  and {{UV}} to {{IR}} few-femtosecond pulses generated by high-energy soliton
  self-compression},\ }\href {https://doi.org/10.1103/PhysRevResearch.2.043037}
  {\bibfield  {journal} {\bibinfo  {journal} {Phys. Rev. Research}\ }\textbf
  {\bibinfo {volume} {2}},\ \bibinfo {pages} {043037} (\bibinfo {year}
  {2020})}\BibitemShut {NoStop}%
\bibitem [{\citenamefont {K{\"o}ttig}\ \emph {et~al.}(2017)\citenamefont
  {K{\"o}ttig}, \citenamefont {Tani}, \citenamefont {Biersach}, \citenamefont
  {Travers},\ and\ \citenamefont {Russell}}]{kottig_generation_2017}%
  \BibitemOpen
  \bibfield  {author} {\bibinfo {author} {\bibfnamefont {F.}~\bibnamefont
  {K{\"o}ttig}}, \bibinfo {author} {\bibfnamefont {F.}~\bibnamefont {Tani}},
  \bibinfo {author} {\bibfnamefont {C.~M.}\ \bibnamefont {Biersach}}, \bibinfo
  {author} {\bibfnamefont {J.~C.}\ \bibnamefont {Travers}},\ and\ \bibinfo
  {author} {\bibfnamefont {P.~S.}\ \bibnamefont {Russell}},\ }\bibfield
  {title} {\bibinfo {title} {Generation of microjoule pulses in the deep
  ultraviolet at megahertz repetition rates},\ }\href
  {https://doi.org/10.1364/OPTICA.4.001272} {\bibfield  {journal} {\bibinfo
  {journal} {Optica}\ }\textbf {\bibinfo {volume} {4}},\ \bibinfo {pages}
  {1272} (\bibinfo {year} {2017})}\BibitemShut {NoStop}%
\bibitem [{\citenamefont {{L{\'o}pez-Zubieta}}\ \emph
  {et~al.}(2018{\natexlab{a}})\citenamefont {{L{\'o}pez-Zubieta}},
  \citenamefont {Jarque}, \citenamefont {Sola},\ and\ \citenamefont
  {Roman}}]{lopez-zubieta_theoretical_2018}%
  \BibitemOpen
  \bibfield  {author} {\bibinfo {author} {\bibfnamefont {B.~A.}\ \bibnamefont
  {{L{\'o}pez-Zubieta}}}, \bibinfo {author} {\bibfnamefont {E.~C.}\
  \bibnamefont {Jarque}}, \bibinfo {author} {\bibfnamefont {{\'I}.~J.}\
  \bibnamefont {Sola}},\ and\ \bibinfo {author} {\bibfnamefont {J.~S.}\
  \bibnamefont {Roman}},\ }\bibfield  {title} {\bibinfo {title} {Theoretical
  analysis of single-cycle self-compression of near infrared pulses using
  high-spatial modes in capillary fibers},\ }\href
  {https://doi.org/10.1364/OE.26.006345} {\bibfield  {journal} {\bibinfo
  {journal} {Opt. Express, OE}\ }\textbf {\bibinfo {volume} {26}},\ \bibinfo
  {pages} {6345} (\bibinfo {year} {2018}{\natexlab{a}})}\BibitemShut {NoStop}%
\bibitem [{\citenamefont {{L{\'o}pez-Zubieta}}\ \emph
  {et~al.}(2018{\natexlab{b}})\citenamefont {{L{\'o}pez-Zubieta}},
  \citenamefont {Jarque}, \citenamefont {Sola},\ and\ \citenamefont
  {Roman}}]{lopez-zubieta_spatiotemporal-dressed_2018}%
  \BibitemOpen
  \bibfield  {author} {\bibinfo {author} {\bibfnamefont {B.~A.}\ \bibnamefont
  {{L{\'o}pez-Zubieta}}}, \bibinfo {author} {\bibfnamefont {E.~C.}\
  \bibnamefont {Jarque}}, \bibinfo {author} {\bibfnamefont {{\'I}.~J.}\
  \bibnamefont {Sola}},\ and\ \bibinfo {author} {\bibfnamefont {J.~S.}\
  \bibnamefont {Roman}},\ }\bibfield  {title} {\bibinfo {title}
  {Spatiotemporal-dressed optical solitons in hollow-core capillaries},\ }\href
  {https://doi.org/10.1364/OSAC.1.000930} {\bibfield  {journal} {\bibinfo
  {journal} {OSA Continuum, OSAC}\ }\textbf {\bibinfo {volume} {1}},\ \bibinfo
  {pages} {930} (\bibinfo {year} {2018}{\natexlab{b}})}\BibitemShut {NoStop}%
\bibitem [{\citenamefont {Marcatili}\ and\ \citenamefont
  {Schmeltzer}(1964)}]{marcatili_hollow_1964}%
  \BibitemOpen
  \bibfield  {author} {\bibinfo {author} {\bibfnamefont {E.~A.~J.}\
  \bibnamefont {Marcatili}}\ and\ \bibinfo {author} {\bibfnamefont {R.~A.}\
  \bibnamefont {Schmeltzer}},\ }\bibfield  {title} {\bibinfo {title} {Hollow
  {{Metallic}} and {{Dielectric Waveguides}} for {{Long Distance Optical
  Transmission}} and {{Lasers}}},\ }\href
  {https://doi.org/10.1002/j.1538-7305.1964.tb04108.x} {\bibfield  {journal}
  {\bibinfo  {journal} {Bell System Technical Journal}\ }\textbf {\bibinfo
  {volume} {43}},\ \bibinfo {pages} {1783} (\bibinfo {year}
  {1964})}\BibitemShut {NoStop}%
\bibitem [{\citenamefont {Heyl}\ \emph {et~al.}(2016)\citenamefont {Heyl},
  \citenamefont {{Coudert-Alteirac}}, \citenamefont {Miranda}, \citenamefont
  {Louisy}, \citenamefont {Kovacs}, \citenamefont {Tosa}, \citenamefont
  {Balogh}, \citenamefont {Varj{\'u}}, \citenamefont {L'Huillier},
  \citenamefont {Couairon},\ and\ \citenamefont
  {Arnold}}]{heyl_scale-invariant_2016}%
  \BibitemOpen
  \bibfield  {author} {\bibinfo {author} {\bibfnamefont {C.~M.}\ \bibnamefont
  {Heyl}}, \bibinfo {author} {\bibfnamefont {H.}~\bibnamefont
  {{Coudert-Alteirac}}}, \bibinfo {author} {\bibfnamefont {M.}~\bibnamefont
  {Miranda}}, \bibinfo {author} {\bibfnamefont {M.}~\bibnamefont {Louisy}},
  \bibinfo {author} {\bibfnamefont {K.}~\bibnamefont {Kovacs}}, \bibinfo
  {author} {\bibfnamefont {V.}~\bibnamefont {Tosa}}, \bibinfo {author}
  {\bibfnamefont {E.}~\bibnamefont {Balogh}}, \bibinfo {author} {\bibfnamefont
  {K.}~\bibnamefont {Varj{\'u}}}, \bibinfo {author} {\bibfnamefont
  {A.}~\bibnamefont {L'Huillier}}, \bibinfo {author} {\bibfnamefont
  {A.}~\bibnamefont {Couairon}},\ and\ \bibinfo {author} {\bibfnamefont
  {C.~L.}\ \bibnamefont {Arnold}},\ }\bibfield  {title} {\bibinfo {title}
  {Scale-invariant nonlinear optics in gases},\ }\href
  {https://doi.org/10.1364/OPTICA.3.000075} {\bibfield  {journal} {\bibinfo
  {journal} {Optica}\ }\textbf {\bibinfo {volume} {3}},\ \bibinfo {pages} {75}
  (\bibinfo {year} {2016})}\BibitemShut {NoStop}%
\bibitem [{\citenamefont {Dudley}\ \emph {et~al.}(2006)\citenamefont {Dudley},
  \citenamefont {Genty},\ and\ \citenamefont
  {Coen}}]{dudley_supercontinuum_2006}%
  \BibitemOpen
  \bibfield  {author} {\bibinfo {author} {\bibfnamefont {J.~M.}\ \bibnamefont
  {Dudley}}, \bibinfo {author} {\bibfnamefont {G.}~\bibnamefont {Genty}},\ and\
  \bibinfo {author} {\bibfnamefont {S.}~\bibnamefont {Coen}},\ }\bibfield
  {title} {\bibinfo {title} {Supercontinuum generation in photonic crystal
  fiber},\ }\href {https://doi.org/10.1103/RevModPhys.78.1135} {\bibfield
  {journal} {\bibinfo  {journal} {Reviews of Modern Physics}\ }\textbf
  {\bibinfo {volume} {78}},\ \bibinfo {pages} {1135} (\bibinfo {year}
  {2006})}\BibitemShut {NoStop}%
\bibitem [{\citenamefont {Nagy}\ \emph {et~al.}(2008)\citenamefont {Nagy},
  \citenamefont {Forster},\ and\ \citenamefont {Simon}}]{nagy_flexible_2008}%
  \BibitemOpen
  \bibfield  {author} {\bibinfo {author} {\bibfnamefont {T.}~\bibnamefont
  {Nagy}}, \bibinfo {author} {\bibfnamefont {M.}~\bibnamefont {Forster}},\ and\
  \bibinfo {author} {\bibfnamefont {P.}~\bibnamefont {Simon}},\ }\bibfield
  {title} {\bibinfo {title} {Flexible hollow fiber for pulse compressors},\
  }\href {https://doi.org/10.1364/AO.47.003264} {\bibfield  {journal} {\bibinfo
   {journal} {Applied Optics}\ }\textbf {\bibinfo {volume} {47}},\ \bibinfo
  {pages} {3264} (\bibinfo {year} {2008})}\BibitemShut {NoStop}%
\bibitem [{\citenamefont {Brahms}\ \emph {et~al.}(2019)\citenamefont {Brahms},
  \citenamefont {Grigorova}, \citenamefont {Belli},\ and\ \citenamefont
  {Travers}}]{brahms_high-energy_2019}%
  \BibitemOpen
  \bibfield  {author} {\bibinfo {author} {\bibfnamefont {C.}~\bibnamefont
  {Brahms}}, \bibinfo {author} {\bibfnamefont {T.}~\bibnamefont {Grigorova}},
  \bibinfo {author} {\bibfnamefont {F.}~\bibnamefont {Belli}},\ and\ \bibinfo
  {author} {\bibfnamefont {J.~C.}\ \bibnamefont {Travers}},\ }\bibfield
  {title} {\bibinfo {title} {High-energy ultraviolet dispersive-wave emission
  in compact hollow capillary systems},\ }\href
  {https://doi.org/10.1364/OL.44.002990} {\bibfield  {journal} {\bibinfo
  {journal} {Optics Letters}\ }\textbf {\bibinfo {volume} {44}},\ \bibinfo
  {pages} {2990} (\bibinfo {year} {2019})}\BibitemShut {NoStop}%
\bibitem [{\citenamefont {Brahms}\ and\ \citenamefont
  {Travers}(2021)}]{brahms_lunajl_2021}%
  \BibitemOpen
  \bibfield  {author} {\bibinfo {author} {\bibfnamefont {C.}~\bibnamefont
  {Brahms}}\ and\ \bibinfo {author} {\bibfnamefont {J.~C.}\ \bibnamefont
  {Travers}},\ }\href {https://doi.org/10.5281/zenodo.5513570} {\bibinfo
  {title} {Luna.jl: A flexible nonlinear optical pulse propagator}},\ \bibinfo
  {howpublished} {https://github.com/LupoLab/Luna.jl} (\bibinfo {year}
  {2021})\BibitemShut {NoStop}%
\end{thebibliography}%
\end{document}